\documentclass[12pt,a4paper]{article}
\usepackage[latin1]{inputenc}
\usepackage{amsmath}
\usepackage{amsfonts}
\usepackage{amssymb}
\usepackage{graphicx}
\newcommand{\dd}{\textrm d}
\newcommand{\etal}{{\it et al. }}
\newcommand{\ie}{{\it i. e. }}

\newcommand{\beq}{\begin{equation}}
\newcommand{\eeq}{\end{equation}}

\newcommand{\vect}[1]{\mathbf{#1}}

\begin{document}
\author{Doru Constantin\footnote{doru.constantin@u-psud.fr}}
\date{Universit\'e Paris-Saclay, CNRS, Laboratoire de Physique des Solides, 91405 Orsay, France}
\title{The Unreasonable Effectiveness of the Nallet Model}
\maketitle
\begin{abstract}
In 1993, Nallet, Laversanne and Roux put forward a simplified model for the intensity scattered by lamellar phases, which was nonetheless very successfully used in fitting experimental results, especially those obtained with powder systems. I argue that the success as well as the simple expression of the model result from an approximate integration over sample orientation, resulting from an implicit integration over the scattering vector component normal to the director.
\end{abstract}

\section{Introduction}

A rigorous description of the intensity scattered by lamellar phases leads to very involved expressions. Nevertheless, In 1993 Nallet, Laversanne and Roux put forward a simplified model \cite{Nallet:1993} (usually known as the ``Nallet model''). Aside from its very tractable expression, this model surprisingly exhibits the power-law exponents associated with an orientationally averaged system (powder sample) although it is derived by a purely one-dimensional approach. This feature, though not explained in the original paper, is probably the main reason of the model's popularity with experimentalists.

I present here an interpretation of this model that is quite different from that of its authors but that hopefully explains better some of its features and its empirical success.


\section{Formalism}

We will write the structure factor in a slightly more general form than in the original paper \cite[Eq. (9)]{Nallet:1993}:
\begin{equation}
\label{eq:SNallet}
S_{\text{Nallet}}(q_z) = \sum_{m,n = 1}^N \mathrm{e}^{-iq_zd(m-n)} \mathrm{e}^{-\frac{1}{2}q_z^2 \left\langle \left( u_m - u_n\right) ^2\right\rangle } 
\end{equation}
\noindent where the argument $q_z$ emphasizes that there is no dependence on the transverse component $\vect{q}_{\bot}$. Equation \eqref{eq:SNallet} reduces to the form given by Nallet \etal in the case of translation invariance along $z$ (\ie when $\left\langle \left( u_m - u_n\right) ^2\right\rangle$ only depends on $m-n$).

Let us define the correlation function $g_{mn}(\vect{r}_{\bot})=\left \langle \left ( u_m(\vect{r}_{\bot})- u_n({\mathbf 0}) \right ) ^2 \right
\rangle$. For finite displacements $\langle u_m^2 \rangle < \infty \, , \, \forall m$:
\begin{equation}
\label{eq:gmnr}
g_{mn}(\vect{r}_{\bot})= \langle u_m^2 \rangle+\langle u_n^2 \rangle-2\langle u_m(\vect{r}_{\bot})u_n({\mathbf 0}) \rangle
\end{equation}
\noindent In the bulk, where $g$ only depends on the difference $m-n$ (and the individual displacements $\langle u_m^2 \rangle$ diverge) the explicit form for the correlation function was obtained in Caill\'{e}'s seminal paper \cite{Caille:1972,deJeu:2003}:
\begin{equation}
\label{eq:gCaille}
g_{mn}(\vect{r}_{\bot})=g_{|m-n|}(r)= \eta \left( \frac{d}{\pi}\right)^2 \left[ \gamma + \frac{1}{2} \text{E}_1 \left( \frac{r^2}{4 \lambda d \, |n-m|}\right) - \ln \left( \frac{2d}{r}\right) \right] 
\end{equation}
\noindent where we invoked the in-plane isotropy of the system to say that $g$ only depends on the absolute value $r  = \left| \vect{r}_{\bot} \right| $. Here, $\lambda = \sqrt{K/B}$ is the penetration length, $\eta = \frac{\pi}{2} \frac{k_BT}{B \lambda d^2}$ is the Caill\'{e} exponent, $\gamma$ is Euler's constant and $\text{E}_1$ the exponential integral.

At this point, let us also introduce the auxiliary function
\begin{equation}
\label{eq:fmndef}
f_{mn}(r,q_z) = \exp \left[ - \frac{q_z^2}{2} g_{mn}(r) \right] 
\end{equation}

The Nallet model only uses the ``vertical'' correlation function:
\begin{equation}
\label{eq:gNallet}
g_{mn}(\vect{r}_{\bot} = \vect{0}) = \left\lbrace \begin{array}{ll}
\eta \dfrac{d^2}{8} \, |m-n|^2 &, |m-n| \text{ small} \\
\eta \dfrac{d^2}{2 \pi ^2} \left[ \ln \left( \pi |m-n|\right) + \gamma \right]  &, |m-n| \gg 1
\end{array}
\right. 
\end{equation}
\noindent see Eq.~(5) in \cite{Nallet:1993} and the discussion immediately above it. In the above notations, the structure factor becomes:
\begin{equation}
\label{eq:SNalletbis}
S_{\text{Nallet}}(q_z) = \sum_{m,n = 1}^N \mathrm{e}^{-iq_zd(m-n)}  f_{mn}(0,q_z)
\end{equation}

\noindent while the full structure factor \cite{deJeu:2003} writes
\begin{equation}
\label{eq:Sqtot}
S(\vect{q}_{\bot}, q_z) = \sum_{m,n = 1}^N \mathrm{e}^{-iq_zd(m-n)}
\int \dd \vect{r}_{\bot} \, \mathrm{e}^{-i {\mathbf q}_{\bot}\vect{r}_{\bot}} \,
\overbrace{\mathrm{e}^{-\frac{1}{2}q_z^2g_{mn}(r)}}^{f_{mn}(r,q_z)} = \sum_{m,n = 1}^N \mathrm{e}^{-iq_zd(m-n)} \tilde{f}_{mn}(q_{\bot}, q_z)
\end{equation}
\noindent by recognizing that the integral in \eqref{eq:Sqtot} is the definition of $\tilde{f}(\vect{q}_{\bot}, q_z)$, the Fourier transform of $f(\vect{r},q_z)$ with respect to $\vect{r}$.

Let us now integrate $S(\vect{q}_{\bot}, q_z)$ over the plane normal to the director:
\begin{equation}
	\label{eq:Sqint}
\begin{split}
	S_{\text{int}}(q_z) &= \int \dd \vect{q}_{\bot} \, S(\vect{q}_{\bot}, q_z) = \sum_{m,n = 1}^N \mathrm{e}^{-iq_zd(m-n)}
	\int \dd \vect{r}_{\bot} \, \overbrace{\int \dd \vect{q}_{\bot} \, \mathrm{e}^{-i {\mathbf q}_{\bot}\vect{r}_{\bot}}}^{\sim \delta(\vect{r}_{\bot})} \, f_{mn}(r,q_z) \\
	&\sim \sum_{m,n = 1}^N \mathrm{e}^{-iq_zd(m-n)} f_{mn}(0,q_z)
\end{split}
\end{equation}
\noindent and we retrieve $S_{\text{Nallet}}(q_z)$ as given in \eqref{eq:SNalletbis}.

\section{Interpretation}

\begin{figure}[htbp]
	\centerline{\includegraphics[width=0.5\textwidth]{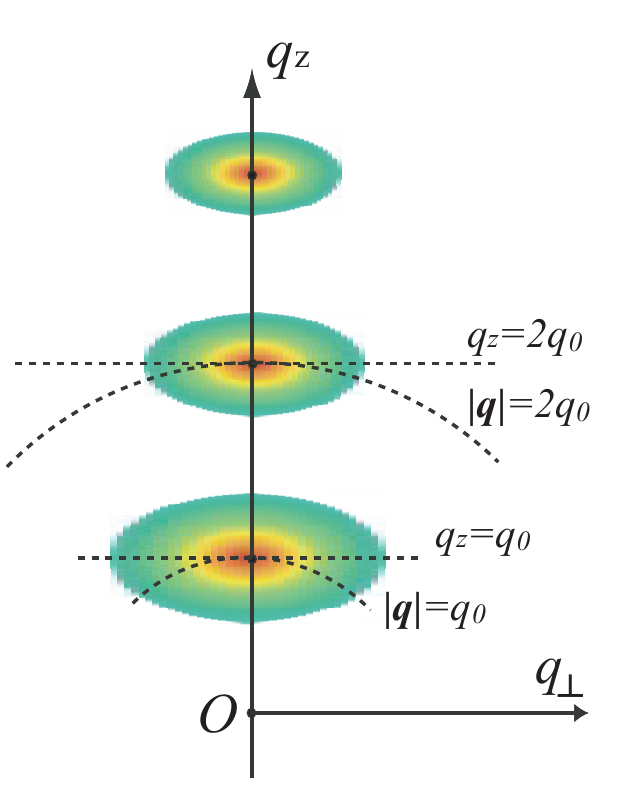}}
	\caption{Typical structure for the scattering from a lamellar phase. The signal is concentrated into ``Bragg sheets'' centered on the $q_z$ axis, at integer multiples of $q_0 = 2 \pi /d$. The two types of integration discussed in the text correspond to the two families of paths shown as dashed line: planes with constant $q_z$ and spheres with constant $\left | \vect{q} \right |$. \label{fig:Scattering}}
\end{figure}

Integrating over $\vect{q}_{\bot}$ at constant $q_z$ is not a very common operation for SAXS users, but ``powder averaging'' (integrating over all sample orientations at constant $q = \left | \vect{q} \right |$) is. In the particular case of lamellar phases, these operations are fairly close, as one can see in Figure~\ref{fig:Scattering}.

For these systems, the scattered signal is concentrated into ``Bragg sheets'' centered on the $q_z$ axis and decays very rapidly with $q_{\bot}$: $S(q_{\bot},q_z = k q_0) \sim q_{\bot}^{-4+2\eta k^2}$ \cite{deJeu:2003}, such that the main contribution comes from the vicinity of the $q_z$ axis, where the planes with constant $q_z$ and the spheres with constant $\left | \vect{q} \right |$ are very close.

An indirect confirmation of the resemblance between $S_{\text{Nallet}}$ and the orientationally integrated signal can be obtained by considering the power-law behavior close to the Bragg ``peaks''. It is known that $S(q_{\bot}=0,q_z = k q_0+\delta q) \sim |\delta q|^{-X}$, with exponents
$X = 2 - \eta k^2$ for aligned samples  and $X = 1 - \eta k^2$ for a powder average \cite{Roux:1988,Kaganer:1991}. Although obtained for a perfectly aligned system, $S_{\text{Nallet}}(q_z)$ exhibits $X = 1 - \eta k^2$ variation around its maxima \cite{Nallet:1993}. The similarity with powder samples was noted by the authors, but with no further explanation. From the discussion above, we can understand this similarity as due to the implicit orientation average performed in going from \eqref{eq:Sqtot} to \eqref{eq:Sqint}.

%
%
%
%
%
%
%



\end{document}